\documentstyle[twocolumn,prl,aps]{revtex}
\tighten
\begin{document}
\draft
\title{Shape of a liquid front upon dewetting}
\author{R. Seemann, S. Herminghaus and K. Jacobs}
\address{Dept. of Applied Physics, University of Ulm, D-89069 Ulm,
Germany} \maketitle
\begin{abstract}
We examine the profile of a liquid front of a film that is
dewetting a solid substrate. Since volume is conserved, the
material that once covered the substrate is accumulated in a rim
close to the three phase contact line. Theoretically, such a
profile of a Newtonian liquid resembles an exponentially decaying
harmonic oscillation that relaxes into the prepared film
thickness. For the first time, we were able to observe this
behavior experimentally. A non-Newtonian liquid - a polymer melt -
however, behaves differently. Here, viscoelastic properties come
into play. We will demonstrate that by analyzing the shape of the
rim profile. On a nm scale, we gain access to the rheology of a
non-Newtonian liquid.

\end{abstract}

The interplay of viscous and viscoelastic properties of a fluid is
of enormous technical importance. The rheology e.g. of a coating
will be adjusted such that standard coating techniques may be
applied. In those thin films, however, the typical rheological
analysis techniques fail. In this Letter we will propose a method
that can even be applied to nm thin films. We will moreover
demonstrate the sensitivity of a liquid profile even to slight
changes in the viscoelastic properties of the material. To the
best of our knowledge, the analysis of a rim profile was only
subject to theoretical studies \cite{1,2,3,12,Bahr}, yet
experimental studies concentrated on the characterization of the
dynamics of dewetting
\cite{Stange,HerminghausSci,Reiter00,7,Jop2}, but not on the shape
of the rim.

A liquid film on a non-wettable substrate is not stable and will
bead up: circular `dry' regions (`holes') appear, the radius of
which grows steadily with time \cite{Reiter92}. If mass
conservation is valid, the material once at the inner part of the
hole will be accumulated in a rim along the perimeter of the hole.
Close to the three phase contact line of the liquid, i.e. on the
`dry' side of the rim, the rim shape is dominated by the receeding
contact angle $\theta$ of the liquid on the surface. On the `wet'
side of the rim, where the actual contact angle is zero, the
liquid merges somehow or other from the crest into the initial
film thickness. From a theoretical point of view, the rim is a
disturbance in the liquid and is expected to decay via an
undulation, the wavelength of which is dominated by the surface
tension of the liquid. To the best of our knowledge, however, this
peculiar behavior was never reported on dewetting experiments. In
our experimental system, short chain polystyrene (PS) on silicon
wafers, an undulatory behavior is observed and measured for the
first time. It is one of the aims of this Letter to explain why
this behavior is only rarely found in experiments and how it can
be induced. Moreover, as we will show in this study, the profile
of the `wet' side of the rim may serve as a fingerprint of the
viscoelastic properties of the liquid.

Using thin atactic PS films on Si wafers as a model system for
liquid front studies has several advantages: the PS melt has a
very low vapor pressure, hence mass conservation is valid, its
viscosity can be tuned by choosing various chain lengths and/or
annealing temperatures, its physical properties are well-known and
PS is available in high purity and low polydispersity. Moreover,
PS films can be studied in the liquid state by atomic force
microscopy (AFM). Polished Si wafers are ideal model substrates
due to their very low surface roughness (rms values below 0.1 nm).
The native oxide layer of the Si wafers renders the surface quite
inert in air, and chemical homogeneity of the surface can be
achieved by well-known cleaning procedures. Moreover, to check the
influence of contact angle on the profile of the liquid front, the
wettability of the Si wafer can be changed by the preparation of a
self-assembled monolayer of octadecyl-trichlorosilane (OTS) on the
surface.

Two types of substrates were used: Type A are Si wafers with a
thin natural oxide layer of 1.6(2) nm \cite{error} (Wacker
Chemitronics, Burghausen, Germany; crystal orientation (100),
p-(boron-)doped, 6-12 $\Omega$cm) on top of which an OTS monolayer
was prepared following standard procedures. Type B substrates are
Si wafers with a thick oxide layer of 191(1) nm (Silchem GmbH,
Freiberg, Germany; (100)-oriented, p-(boron-)doped, conductivity
$>$ 1 $\Omega$cm).

Thin polystyrene films (molecular weight between 2.05 kg/mol and
600 kg/mol, M$_{w}$/M$_{n}$  = 1.02 - 1.05, purchased from PSS,
Mainz, Germany and Polymer Labs, Church Stratton, UK) were spin
cast from toluene solution onto the substrates. Before coating,
type B substrates, cut in ca. 1 cm$^{2}$ pieces, were freed from
Si dust by treatment with a CO$_{2}$-jet \cite{17}. The samples
were then `degreased' by sonication in ethanol, acetone and
toluene. Residual hydrocarbons were removed by immersing in a
fresh 1:1 H$_{2}$O$_{2}$ (conc.)/H$_{2}$O$_{2}$ (30\%) solution
for 30 min, followed by a thorough rinse in hot Millipore$^{TM}$
water. Due to the large contact angle of the toluene solution on
type A substrates, PS films were first spin coated onto freshly
cleaved mica surfaces (B $\&$ M  Mica Co., Flushing, NY), then
floated onto Millipore$^{TM}$ water and picked up by the type A
substrates.

Cleaning and coating was performed in a class 100 clean room
atmosphere. The thickness of the Si wafer's oxide layer and of the
polymer films were measured by ellipsometry (Multiscope by Optrel
GdBR, Berlin, Germany) at various angles of incidence. Further
characterization of wafers and polymer films was done by atomic
force microscopy (AFM) (Multimode III by Digital Instruments,
Santa Barbara, CA) in Tapping Mode$^{TM}$. Annealing took place on
a temperature-controlled hot plate in a class 100 clean room
atmosphere or in situ on top of the AFM sample holder.

The inset of Fig. 1 presents an AFM scan of a typical hole in a
6.6(2) nm \cite{error} thick PS(2.24k) film on a wafer B. The
diagram shows a radial cross section of this hole. Most
outstanding feature of the cross section is the way the rim decays
into the unperturbed film. The rim profile is asymmetric, with
higher slopes near the three-phase contact line and a `trough' on
the wet side of the rim where it meets the undisturbed film. Here,
the form resembles a damped harmonic oscillation. We describe the
profile by the contact angle $\Theta$ of the liquid at the three
phase contact line, the height $A$ of the rim at its crest, the
depth $B$ of the first trough and the slope $\tan \alpha$ of the
rim profile on the wet side at the level of the initial film
thickness $h_{0}$, as depicted in Fig. 1. The trough is also
visible in the top view AFM scan in the form of a dark circle
surrounding the wet side of the rim (inset of Fig. 1).

From the temporal evolution of the experimental rim profile as
shown in Fig. 2, we can determine the width $a$ and height $A$ of
the rim, the depth $B$ and the width $b$ of the trough, and the
slope $\tan \alpha $ as a function of time. Moreover, by in situ
AFM scanning, we can record the dewetting velocity $v$. Within the
experimental error, the contact angle $\Theta$ stays constant
during hole growth, which is consistent with other experimental
and theoretical studies \cite{2,13}. Fig. 2b) and c) show how
$|B|$ and $\alpha$ vary with increasing height $A$ of the rim.
From a certain value $A_{exp}^{\star}$ of the rim height onwards,
which will call the late stage in the following, both $B$ and
$\alpha$ reach a plateau value. Below $A_{exp}^{\star}$, in our
terms now the early stage, $|B|$ as well as $\alpha$ increase -
within the experimental error - linearly with $A$.

For a physical understanding of the occurrence and the temporal
evolution of the trough we analytically solve in the following the
linearized thin film equation in the lubrication approximation
\cite{18,19,20}:

\begin{equation}
\frac{\partial h}{\partial t} + \frac{\sigma h_{0}^{3}}{3 \eta}
\frac{\partial ^4  h}{\partial x ^4}  -v \frac{\partial
h}{\partial x} = 0 \; , \label{6}
\end{equation} where $h$ is the film thickness, $\sigma$ and $\eta$
are the surface tension and viscosity, and $v$ the velocity of the
three phase contact line. We hereby neglect the disjoining
pressure  -$\Phi'(h)$, since we know from an earlier study that
its impact on the dewetting morphology is negligible as compared
to the interfacial forces involved here \cite{7,Phi}. The
`traveling wave' solution $ h(\xi)= h(x - vt)$ of Eq. (\ref{6})
reads
\begin{equation}
h(\xi) = \exp \left( - \frac{2 \pi}{\sqrt{3} \lambda } \: \xi
\right) \cdot \cos \left( \frac{2 \pi}{\lambda } \: \xi
 \right)   \; ,    \label{9}
\end{equation}
where the wavelength $\lambda$ is defined as
\begin{equation}
\lambda := \frac{4 \pi}{h_{0} \sqrt{3}} \cdot \sqrt[3]
{\frac{\sigma}{3 \eta v}} \; . \label{10}
\end{equation}
Here, $\eta$ denotes the viscosity of the polymer. Eq. (\ref{9})
represents a damped harmonic oscillation. The ratio of $h(\xi)$ at
two consecutive extrema is:

\begin{equation}
\left| \frac{B}{A} \right| =  \exp  \left(- \frac{\pi}{\sqrt{3}}
\right)  \approx 0.163 \label{12}
\end{equation}
and is plotted as the dotted line in Fig. 3. This ratio is
independent of any experimental parameter of the system, in
particular of the dewetting velocity. It is universal for any rim
profile of a retracting Newtonian liquid if the film thickness is
small as compared to the capillary length. Although, it cannot be
valid for the entire process of dewetting. Due to accumulation of
material inside the rim, its width $a$ will be much larger than
the width $b$ of the trough. In other words, as soon as the crest
of the rim exceeds a certain height $A_{theo}^{\star}$, the
description of the entire rim profile as a damped harmonic
oscillation with a single wavelength $\lambda$=$2b$ in Eq.
(\ref{10}) will be no more adequate.

In Fig. 3, the experimental data of $|B$/$A|$ are plotted over
$A$. They reveal that - within the experimental error -  $B/A$ =
const. for $A < A_{exp}^{\star}$. Yet the constant is roughly a
factor of two smaller than 0.163. For $A > A_{exp}^{\star}$, $B/A$
decreases with increasing $A$.

In order to extend the theoretical description to the late stage
of dewetting, we have to introduce a boundary condition. From
experiment, we have learned that $B$ as well as $\alpha$ are
constant in the late stage, c.f. Fig. 2b and c.  For
$A~>~A^{\star}$ we therefore introduce into the calculation the
experimental plateau value $\alpha$ = 1.0(1)$^{\circ}$ as a
boundary condition. The trough depth $B$ hence is simply obtained
by identifying in Eq. (\ref{9}) the slope of $h(\xi)$ at its first
zero as $\tan \alpha$. So, $B$ is a function of the experimental
parameters $\sigma$, $\eta$, $h_{0}$ and dewetting velocity $v$.

Figure 3 shows the result for $|B/A|$ obtained with the boundary
condition $\alpha=$1.0(1)$^{\circ}$ for $A$ $>$ $A_{theo}^{\star}$
(dashed line). Here, the trough depth $B$ is calculated with
$\sigma$ = 30.8 mN/m, $\eta$ = 300 Pa$\cdot$s (Ref. \cite{9}),
$h_{0}$ = 11 nm, and $v(t)$ taken from the online AFM measurement
shown in Fig. 2 \cite{Vau}. The point of intersection of dashed
and dotted line at $A$=$A_{theo}^{\star}$ corresponds to the value
of $A_{exp}^{\star}$ at which the experimental values of $|B/A|$
are starting to decrease. This point hence is name $A^{\star}$ in
the following. For $A$ $>$ $A^{\star}$, the course of the
theoretical curve for $|B/A|$ resembles that of the experimental
data, but the absolute values differ by a factor of about 1.5.

To sum up, one may say that the theoretical curves presented above
well describe the qualitative development of the height of the rim
with respect to the depth of the trough. A quantitative agreement,
however, is not reached yet. In fact, we found that the trough may
even be completely suppressed by increasing the molecular weight
of the polymer. In Fig.~4 we show profiles of holes in PS films on
type A substrates of about the same film thickness (50 nm) and
hole radius, yet of various molecular weight: 2 kg/mol, 18 kg/mol,
101 kg/mol and 600 kg/mol. The most outstanding feature in these
profiles is that the trough vanishes and the width of the rim
grows with increasing molecular weight of the polymer.

It is interesting to note that the PS(2k) and the PS(18k) profiles
- both polymers are below the entanglement length \cite{Elias}) -
look quite similar, yet the change in viscosity is about four
orders of magnitude \cite{eta}. The other two profiles, however,
greatly differ in shape, yet the increase in viscosity as compared
to PS(18k) is only two (four) orders of magnitude for PS(101k)
(PS(600k)). These results suggest that the alteration in profile
shape is caused by an increase in viscoelastic properties, i.e.,
the ratio of viscosity and elastic modulus at the typical shear
rates to which the material is subject upon dewetting.

To corroborate this hypothesis, we performed dewetting experiments
with an entangled polymer melt of a single chain length and varied
the temperature. This is expected to change the viscoelastic
response in a way completely different from changing the molecular
weight. We prepared 50(2) nm thick PS(65k) films on type A wafers.
The samples were heated up to different temperatures for a certain
time, namely until the holes had reached a radius of about 5
$\mu$m. That way, we changed the viscosity from $\eta$=
4$\cdot10^{4}$ Pa$\cdot$s (T=145$^{\circ}$C) to $\eta$=
5$\cdot10^{6}$ Pa$\cdot$s (T=125$^{\circ}$C). The profiles are
shown in the inset of Fig. 4. We have normalized the data to
account for the slightly different hole sizes. As can be seen in
the inset, the profiles are similar, as opposed to the large
figure.

Since the substrate is identical for all polymer films shown, the
friction at the interface between melt and substrate remains
unchanged for all polymer films. The only difference is that with
increasing chain length the \textit{viscoelasticity} of the
polymer melt plays a more and more pronounced role \cite{4}. In a
theoretical calculation of the stability condition of thin
polymeric films, Safran and Klein \cite{21} found that
viscoelasticity stabilizes a liquid film against surface
undulations, which corroborates our results.

In summary we may say that we report for the first time on an
undulatory behavior of the rim of a dewetting polymer film. Most
experiments done before with dewetting films were performed on
polymers of much higher molecular weight than 2~kg/mol. Hence, the
viscoelastic properties of the melt have suppressed the
undulation.  We introduced a simple model that is able to explain
the qualitative behavior of rim and trough of a short-chain
polymers melt. A  quantitative comparison of theoretical and
experimental results reveals a difference of roughly a factor of
2. For long-chain polymers, the model fails completely due to the
absence of the trough. Analyzing the reason for the factor 2 or
for the complete failure of the model, respectively, we came to
the conclusion that the viscoelastic properties of the polymer
melt are to blame. In is interesting to note here that on the rim
profiles an influence of the elastic properties of the melt can be
found for chain lengths at which bulk rheometers detect purely
viscous properties \cite{4}. The rim profile therefore can be
viewed as an extremely sensitive rheometer for small volumes on nm
scales.

\acknowledgements This work was funded by the German Science
Foundation under grant number JA905/1. We also acknowledge
generous support of Si wafers by Wacker, Burghausen, Germany.

\begin{figure}
\caption{Inset: AFM scan of a hole in a 6.6(2) nm thick PS(2.24k)
film on a wafer B, 30 min at T = 80 $^{\circ}$C. The large diagram
shows a radial cross section of this hole, where $A$ describes the
height of the rim, $B$ the depth of depression and $L$ the length
from the maximum of the rim to the point where the rim crosses the
level of the unperturbed film.}

\caption{a) Cross sections of a growing hole in a 11.0(4) nm thick
PS(2.05k) film on a type B wafer (191.2(2) nm thick oxide layer)
at a temperature of T = 90$^{\circ}$C, as gained from an in situ
AFM scan. Note that the profiles have been shifted laterally such
that the three phase contact `points' fall onto one single point.
The maximum slope at the level of the unperturbed film thickness
is $\tan \alpha=0.026$. b) Depth of depression $B$ over height of
rim $A$ of the hole and c) `wet' contact angle $\alpha$ plotted
over height of rim $A$ of the hole.}

\caption{Ratio ($B/A$) over rim height $A$. Dots: experimental
data for a 11.0(4) nm thick PS(2.05k) film on a type B wafer at a
temperature of T=90$^{\circ}$C. Lines: theoretical expectation in
lubrication approximation for the same system gained from the
analytic calculation without any elastic properties. Dotted line:
solution for the early stage $B/A$ $\approx$ 0.163. Dashed line:
solution for the later stage using reduced viscosity $\eta$ = 300
Pas with the boundary condition gained from experiment:
$\alpha_{theo}$ = 1.0$^{\circ}$.}

\caption{Cross sections as gained from AFM scans of holes in
polystyrene films with different chain length on wafers of type B:
solid line: PS(2k), dashed line: PS(18k), dot-dashed line:
PS(101k), dotted line: PS(600k). The thickness $h$ of the films
and the diameters of the holes are chosen to be about the same.
Inset: Hole profiles of PS(65k) taken a temperatures between 125 -
145~$^{\circ}$C. Differences in film thickness and hole size lead
to slight differences of the profiles.}
\end{figure}


\begin{references}
\bibitem{Phi}Since we also find an undulative behavior also in
60~nm thick PS(18k) films, the influence of $\phi$ may indeed be
neglected
\bibitem{Vau}For very small radii, the velocity of the contact line behaves
like $v \propto \sqrt[3]{1/t}$ passing into a $v = const.$
behavior for very large radii \cite{5,12}. A similar result is
given by the simulation in Ref. \cite{2}. This is important due to
the fact that the calculated depth of depression $|B|$ increases
with decreasing $v \cdot \eta$. Hence, if $v$ reaches a constant
value, $|B|$ reaches a maximum value ($|B|<h_{0}$) and will not
lead to an undulation-induced break-up of the film as it was
speculated in Ref. \cite{1}, even for very long times.
\end{references}
\end{document}